\documentclass[aps,prd,amsmath,amsfonts,amssymb,eqsecnum,nofootinbib,
floatfix,secnumarabic,preprintnumbers,nobalancelastpage,
onecolumn,notitlepage]{revtex4-1}
\usepackage{amsmath,amsfonts,amssymb,xcolor,graphicx,hyperref}
\usepackage[utf8x]{inputenc}
\usepackage[section]{placeins}
\usepackage{cancel}
\usepackage{subfigure,color,dsfont,slashed,multirow}
\usepackage{fontenc}
\usepackage{pbox}
\usepackage{cancel}
\usepackage{setspace}
\usepackage{float}
\usepackage{mathtools,nccmath}%
\usepackage{etoolbox, xparse} 
\usepackage{array}
\usepackage[normalem]{ulem}
\usepackage{amstext}
\usepackage{soul}
\usepackage{cleveref}

\DeclarePairedDelimiterX{\abs}[1]\lvert\rvert{\ifblank{#1}{\,\cdot\,}{#1}}

\let\oldabs\abs
\def\abs{\futurelet\testchar\MaybeOptArgAbs}
\def\MaybeOptArgAbs{\ifx[\testchar\let\next\OptArgAbs
	\else \let\next\NoOptArgAbs\fi \next}
\def\OptArgAbs[#1]#2{\oldabs[#1]{#2}}
\def\NoOptArgAbs#1{\ifblank{#1}{\oldabs{}}{\oldabs[\big]{#1}}}

\DeclarePairedDelimiterX{\set}[1]\{\}{\setargs{#1}}
\NewDocumentCommand{\setargs}{>{\SplitArgument{1}{;}}m}
{\setargsaux#1}
\NewDocumentCommand{\setargsaux}{mm}
{\IfNoValueTF{#2}{#1}{\nonscript\,#1\nonscript\;\delimsize\vert\nonscript\:\allowbreak #2\nonscript\,}}
\let\oldset\set
\def\set{\futurelet\testchar\MaybeOptArgSet}
\def\MaybeOptArgSet{\ifx[\testchar \let\next\OptArgSet
	\else \let\next\NoOptArgSet \fi \next}
\def\OptArgSet[#1]#2{\oldset[#1]{#2}}
\def\NoOptArgSet#1{\OptArgSet[\big]{#1}}
\def\lsim{\raise0.3ex\hbox{$\;<$\kern-0.75em\raise-1.1ex\hbox{$\sim\;$}}}
\def\gsim{\raise0.3ex\hbox{$\;>$\kern-0.75em\raise-1.1ex\hbox{$\sim\;$}}}

\hypersetup{colorlinks=true,linkcolor=blue,citecolor=magenta,filecolor=magenta,
	urlcolor=cyan}
\newcolumntype{P}[1]{>{\centering\arraybackslash}p{#1}}

\hypersetup{bookmarks=true,unicode=true,pdftoolbar=true,pdfmenubar=true,
	pdffitwindow=false,pdfstartview={FitH},pdftitle={NonUniUMSSM},
	pdfauthor={Ya\c{s}ar Hi\c{c}y\i lmaz,Stefano Moretti,Levent Solmaz}, pdfsubject={Subject},
	pdfcreator={Creator},pdfproducer={Producer}, pdfkeywords={Non-minimal 
		supersymmetry}{Beyond the Standard Model Physics}{SUSY},
	pdfnewwindow=true,colorlinks=true,
	linkcolor=blue,citecolor=magenta,filecolor=magenta,urlcolor=cyan}
\newcommand{\be}{\begin{equation}}
	\newcommand{\ee}{\end{equation}}
\def\bsp#1\esp{\begin{split}#1\end{split}}
\renewcommand{\figureautorefname}{Fig.}

\def\sectionautorefname~#1\null{Sec.~(#1)\null}
\def\subsectionautorefname~#1\null{sub--Sec.~(#1)\null}
\def\figureautorefname~#1\null{Fig.~#1\null}
\def\tableautorefname~#1\null{Table~#1\null}
\def\equationautorefname~#1\null{Eq.~#1\null}

\newcommand{\mg}{\textsc{MG5\_aMC}}

\newcommand{\mo}{\textsc{MicrOMEGAs}}

\newcommand{\sa}{\textsc{SARAH}}
\newcommand{\spheno}{\textsc{SPheno}}

\mathchardef\mhyphen="2D

\newcommand{\beq}{\begin{equation}}
	\newcommand{\eeq}{\end{equation}}
\newcommand{\bea}{\begin{eqnarray}}
	\newcommand{\eea}{\end{eqnarray}}

\begin{document}

{\title{Family Non-Universal U(1)$^\prime$ Model with Minimal Number of Exotics\\}
	
	\author{Ya\c{s}ar Hi\c{c}y\i lmaz$^{1,2}$\footnote{E-mail: yasarhicyilmaz@balikesir.edu.tr}}
	\author{Stefano Moretti$^2$\footnote{Email: s.moretti@soton.ac.uk}}
	\author{Levent Solmaz$^{1}$\footnote{Email: lsolmaz@balikesir.edu.tr}}
	
    \affiliation{$^1$Department of Physics, Bal\i kesir University, TR10145, Bal\i kesir, Turkey}
	\affiliation{$^2$ School of Physics $\&$ Astronomy, University of Southampton, Highfield, Southampton SO17 1BJ,UK}

	\date{\today}

	\begin{abstract}
{We have studied phenomenological implications of several family non-universal U(1)$^\prime$ sub-models in the  U(1)$^\prime$-extended Minimal Supersysmmetric Standard Model (UMSSM) possesing an extra down quark type exotic field. In doing this, we have started by enforcing  anomaly cancellation criteria to generate a number of solutions in which the extra  U(1)$^\prime$ charges of the particles are treated as free parameters. We have then  imposed existing bounds coming from colliders and astrophysical observations on the assumed sub-models and  observed that current limits dictate certain charge orientations, for instance, $Q_{H_u}\sim Q_{H_d}$ is preferred in general and the charge of the singlet $Q_S$ cannot be very small ($|Q_S|>$ 0.4) even if any of the charges is allowed to take any value within the $[-1, 1]$ range.  We have finally studied the  potential impact of such non-universal charges on $Z'$ mediated processes and made predictions for existing and future experiments. It has turned out that UMSSMs with or without the presence of light  exotic quarks can yield distinguishable signatures if non-universal charges are realised  in the leptonic sector of such models.}
	\end{abstract} 
	
	\maketitle
	
\section{Introduction}
\label{sec:intro}
	
	Due to increasing pressure stemming from experiments performed at the Large Hadron Collider (LHC) on the Minimal Supersymmetric Standard Model (MSSM), that is, the simplest and most popular supersymmetric extension of the Standard Model (SM), alternative supersymmetric scenarios, non-minimal in nature,  are under deep scrutiny. Among these models, UMSSMs are some of the most attractive and best motivated alternatives to the MSSM. Such extensions can stem from superstring \cite{Cvetic:1996mf} and Grand Unified Theories (GUTs) \cite{Hewett:1988xc}. As a matter of fact, recently, these models attracted a great deal of interest in the literature \cite{Barr:1985qs,Cvetic:1995rj,Cleaver:1997nj,Cleaver:1997jb,Ghilencea:2002da,King:2005jy,Diener:2009vq,Langacker:2008yv,Frank:2013yta,Frank:2012ne,Demir:2010is,Athron:2015tsa,Athron:2009bs,Frank:2020pui,Frank:2020kvp}. In addition to inheriting the positive aspects of the MSSM, like resolution to the gauge hierarchy problem \cite{Gildener:1976ai,Weinberg:1978ym,Susskind:1978ms}, stability of the Higgs potential \cite{Degrassi:2012ry,Bezrukov:2012sa,R.:2019ply}, unification of the gauge couplings \cite{Georgi:1974sy} and offering viable candidates for Dark Matter (DM), such gauge extended models also offer novel explanations where certain deficiencies of the MSSM can be cured. This of course happens at the cost of facing with the additional parameters. In this direction, for instance, in its Supersymmetric versions, a naturally occurring $\mu$ parameter around the weak scale can be given as a well known example. 
	
	Indeed, on the one hand, having additional particles and parameters diverts one from the minimalism (which is not a standard feature of Nature) while, on the other hand, benefits of possessing additional degrees of freedom can not be overlooked. As is well known, in the MSSM, there is no reason to force the $\mu$ term around the weak scale, the so-called $ \mu $-problem \cite{Bae:2019dgg}. Contrary to the MSSM, in U(1)$^\prime$ extensions of it, an effective $\mu$ term is present as $\mu=\frac{1}{2} \lambda v_s$, where natural choices of  the Vacuum Expectation Value (VEV) of the singlet ($v_s$) and  the Yukawa coupling ($\lambda$)  automatically yield a value for the  problematic $\mu$ around the weak scale \cite{Suematsu:1994qm,Jain:1995cb,Nir:1995bu}. Moreover, since $\mu$ depends on two different parameters, we have an enormous freedom in obtaining  a value of the $\mu$ parameter capable of generating successful Radiative Electro-Weak Symmetry Breaking (REWSB). Of course, for the same value of $\mu$,  different $\lambda$ and $v_s$  choices onset different phenomenologies, which should then be studied in detail within these U(1)$^\prime$ models. This can be very important, for instance, in so-called secluded U(1)$^\prime$ models where one has more degrees of freedom in comparison to the standard U(1)$^\prime$ models, due to additional singlet VEVs \cite{Erler:2002pr, Frank:2020kvp}. {In addition to these motivations, on the experimental side, there are also hints that flavour physics in the SM is not adequate to provide an explanation for the non-universal leptonic decays observed by the LHCb
	Collaboration \cite{Aaij:2014ora}. Since the experimental results on $R_K$ and $R_{K^*}$  show tendencies to deviate from the SM prediction (around $2.5\sigma$), there are many papers where these observations are explained
	with a family non-universal $Z'$ boson. Of course if the universality is broken, numerous alternatives
	appear and the origin of the non-universal couplings should be studied in detail (see, e.g., \cite{King:2018fcg}).}

	{Like in the SM, also in the MSSM quarks and leptons couple universally. In contrast, in U(1)’ extended
	versions of both, non-universal couplings can be accommodated.} Related to U(1)$^\prime$ extensions, it should be noted that, by introducing a new gauge symmetry into the model, the presence of a new massive neutral gauge boson, called $Z'$,  is inevitable and its couplings are determined by the assumed charges for which the Anomaly Cancellation Conditions (ACCs) stemming from the additional U(1)$^\prime$ group insertion should be respected. This is usually done by adding several exotic states to the spectrum \cite{Cheng:1998nb, Erler:2000wu}. However, this is not the only way to satisfy ACCs and one can obtain an anomaly-free supersymmetric U(1)$^\prime$ model without any exotics, called minimal UMSSM, as was shown by \cite{Demir:2005ti}. As can be inferred, one can also select a hybrid approach in which the presence of the exotic states and non-holomorphic in the Lagrangian terms can be traded accordingly for flavour non-universality. This work can be categorised as one  of the simplest examples of such a hybrid approach, in which  we will scrutinise the possibility of non-universal U(1)$^\prime$ charges with an additional exotic quark superfield, $\widehat{D_x}$, for the UMSSM.
	
	Of course, by having non-universal charges one may end up with dangerous flavour violating processes, this is especially true for the quark sector, so we restrict our modifications to the leptonic sector only.  As we will see, the freedom of choosing different U(1)$^\prime$ charges will yield important modifications in certain sectors of the model, for instance, the additional $Z'$ boson may behave very differently  and this is important not only from the perspective of theory but  it also has remarkable implications for experiments. As a matter of fact, ranging from extra dimensions to additional U(1)$^\prime$ extensions there are numerous models which predict the existence of a new gauge boson (the $Z’$) the properties of which depend heavily on the assumptions related with the model construction, like its couplings with ordinary matter fields. This additional heavy gauge boson may be realised in supersymmetric or non-supersymmetric variants of the SM with or without new exotic particles. Hence, probable unusual behaviours of the $Z'$ boson is an interesting subject which can give clues as to the structure of the underlying model. This possibility can be mentioned among one of our main motivations for this study.
	
	Briefly, in this paper, we will study a supersymmetric U(1)’ model with an additional exotic field with non-universal U(1)$^\prime$ charges in the lepton sector. In doing this we first  aim at finding the possible charge configurations which satisfy not only the ACCs but also experimental low energy constraints. This provides us with numerous well-motivated U(1)$^\prime$ sub-models which give new and different signals for the detection of the extra $Z’$ at a time when  no  experimental confirmation of its existence does exist yet. Since its couplings are not known either, we will especially search for non-universal charges (with additional U(1)$^\prime$ charges only in the lepton sector, as intimated) in order to probe how many degrees of freedom actually exist for such a scenario. It is then easy to deduce that the  presence of non-universal couplings in the leptonic sector allows for differing $Z'$ decay rates in comparison to universal charges, so that this is very important for $Z’$ searches, in both setting limits and extracting signals, as all of the latter are obtained room the neutral Drell-Yan (DY) channel. As a by-product of our study we will thus be able to see if a preferred theoretical charge configuration is testable experimentally, in a bottom-up approach that will seek evidence of the new $Z'$ state precisely from its signals involving the other new features of the model, i.e., the coloured 
exotic states and the anomalous leptonic couplings. In fact, we will  investigate the potential signatures of the exotics and the non-universality of  the U(1)$^\prime$ charges at both hadron and lepton colliders, both present and future ones.  
	
	The outline of the rest of the paper is as follows.  In the Section \ref{sec:model} we will introduce the ACCs and other salient features of our model. After summarising our scanning procedure and enforcing experimental constraints in Section \ref{sec:scan}, we present our results over the surviving U(1)$^\prime$ charges and discuss the corresponding particle mass spectrum and $Z'$ production and decay modes in Section \ref{sec:results}. Finally, we summarise and conclude in Section \ref{sec:cons}.

\section{The Model \label{sec:model}}

	In this section we will present the salient features of  our model, which includes an extra Abelian group. In fact, the model extends the MSSM gauge structure with an extra U(1)$^\prime$ symmetry that can arise from any possible string or GUT theory \cite{Langacker:2008yv,Cvetic:1995rj}.
	
	The superpotential in the model allows Yukawa couplings for the quarks and third family leptons as well as  couplings for the exotic fields given by
\begin{eqnarray}
	\widehat{W}&=&h_u\widehat{Q}\cdot \widehat{H}_u \widehat{U}+
	h_d\widehat{Q}\cdot \widehat{H}_d \widehat{D} + h_{\tau}\widehat{L}_3\cdot
	\widehat{H}_d \widehat{E}_3 + \lambda \widehat{S}\widehat{H}_u \cdot
	\widehat{H}_d + {h_{\nu}} \widehat{L}\cdot
	\widehat{H}_u \widehat{N} + {\kappa} \widehat{S} \widehat{D}_x
	\widehat{{\overline{D}}}_x,
	\label{eq:superpot}
\end{eqnarray}
	where $\hat{Q}$ and $\hat{L}_3$ denote the left-handed chiral superfields for the quarks and third family leptons while $\hat{U}$, $\hat{D}$, $\hat{E}_3$ and $\hat{N}$ stand for the right-handed chiral superfields of $u$-type quarks, $d$-type quarks, $ \tau$-type leptons and neutrinos, respectively. Here, $H_{u}$ and $H_{d}$ are the MSSM Higgs doublets and $h_{u,d,\tau}$ are the Yukawa couplings to the matter fields. Then, $h_{\nu}$ is  the Yukawa coupling responsible for generating neutrino masses. Additionally, $\widehat{D}_x$ , $ \widehat{{\overline{D}}}_x$ and $ \widehat{S} $ are   chiral superfields while $ S $ is a singlet under the MSSM group and its VEV, $ \left\langle S\right\rangle =  v_s / \sqrt{2}$, is responsible for the breaking of the U(1)$^\prime$ symmetry. The MSSM bilinear mixing term $\mu H_{d}H_{u}$ is forbidden by the U(1)$^\prime$ invariance and is induced by $ \left\langle S\right\rangle  $ as $ \mu_{\rm eff} = \lambda v_s / \sqrt{2} $ (i.e., an effective $\mu$ term). 
	
	As seen from  Eq. (\ref{eq:superpot}), for the ACCs in the model with family non-universal U(1)$^{\prime}$ charges, some of the Yukawa couplings are forbidden in the superpotential. This results in massless fermions. However, without any anomalies, the non-holomorphic soft supersymmetry breaking terms involving the 'wrong' Higgs field can lead to fermion masses at one loop by gluino or neutralino exchange \cite{Demir:2005ti}. The non-holomorphic terms in this model can be written as follows
	
\begin{equation}
	-\mathcal{L}^{NH}=T_{e}^\prime \tilde{L}_1 H_{u}\tilde{E}_1^{c} +T_{\mu}^\prime \tilde{L}_2 H_{u}\tilde{E_2^{c}}+h.c.
	\label{UMSSM_NH}
\end{equation}

	The fields in  Eqs. (\ref{eq:superpot}) and (\ref{UMSSM_NH}), together with their quantum numbers, are listed in Tab. \ref{tab:charge}. Also, the most general holomorphic Lagrangian responsible for soft supersymmetry breaking with the exotic sector is
	
\begin{align}
	-{\mathcal{L}}_{soft} &= \sum_a M_a \lambda_a\lambda_a-T_{\lambda}\lambda SH_dH_u 
	-T_u h_u U^c Q H_u-T_{d} h_d D^c QH_d-
	T_\tau h_\tau E_3 L_3 H_d-T_{\kappa}\kappa S {D_x {\overline{D}_x}} +h.c.  \nonumber \\
	&+ m_{H_u}^2|H_u|^2+ m_{H_d}^2|H_d|^2+m_S^2|S|^2+ 
	m_{Q}^2 \widetilde Q\widetilde Q
	+m_{U}^2 \widetilde U^c\widetilde U^{c}+m_{D}^2 \widetilde D^c\widetilde D^{c}
	+m_{L}^2 \widetilde L\widetilde L\nonumber \\
	&+ m_{E}^2 \widetilde E^c\widetilde E^{c}+m_{X}^2 \widetilde{D}_x \widetilde{D}_x+m_{\bar{X}}^2 \widetilde{\overline{D}}_x \widetilde{\overline{D}}_x +h.c. \, ,
	\label{eq:soft}
\end{align}
	where $m_{\tilde{Q}}$, $m_{\tilde{U}}$, $m_{\tilde{D}}$, $m_{\tilde{E}}$, $m_{\tilde{L}}$,$m_{H_{u}}$, $m_{H_{d}}$, $m_{\tilde{S}}$, $m_{X}$  and $m_{\bar{X}}$ are the mass matrices of the scalar particles while $M_{a}\equiv M_{1},M_{2},M_{3},M_{4}$ stand for the gaugino masses. Further, $T_{\lambda}$, $T_{u}$, $T_{d}$ , $T_{\tau}$ and $T_{\kappa}$ are the trilinear scalar interaction couplings.

\begin{table}[!t]
		\addtolength{\tabcolsep}{0.001pt}
		\begin{tabular*}{0.99\textwidth}{@{\extracolsep{\fill}} ccccccccccccccccc}
			\hline \hline Gauge group/Field & $\widehat{Q}$ & $\widehat{U}$ &
			$\widehat{D}$ & $\widehat{L}_i$ &$\widehat{N}_i$ & $\widehat{E}_i$ & $\widehat{H}_u$ & $\widehat{H}_d$ & $\widehat{S}$ &  $\widehat{D_x}$ &
			$\widehat{{\overline{D}_x}}$ 
			\\\hline
			$\;$  SU(3)$_C$ & 3 & $\overline{3}$ &  $\overline{3}$ & 1& 1 & 1& 1& 1 & 1& 3 &$\overline{3}$\\
			$\;$  SU(2)$_L$ & 2 & 1 &  1 & 2 &1 &  1& 2& 2 &1 & 1 &1 \\
			$\;$  U(1)$_Y$ & 1/6 & -2/3 &  1/3 & $-1/2$ &0 &  1& 1/2& $-1/2$ &0  &$Y_{D_x}$ &$-Y_{D_x}$ \\
			$\;$  U(1)$^{\prime}$ & $Q_{Q}$ & $Q_{U}$ & $Q_{D}$& $Q_{L_i}$& $Q_{N_i}$& $Q_{E_i}$&$Q_{H_u}$&$Q_{H_d}$& $Q_{S}$ &$Q_{D_x}$&$Q_{{\overline{D}_x}}$  
			\\\hline\hline
		\end{tabular*}
		\caption{\sl\small Gauge quantum numbers of quark ($\widehat{Q},
			\widehat{U}, \widehat{D}$), lepton ($\widehat{L}_i, \widehat{N}_i,
			\widehat{E}_i$), Higgs ($\widehat{H}_u, \widehat{H}_d$) and exotic
			quark ($\widehat{D_x}, \widehat{{\overline{D}_x}}$) superfields. The index i runs over three families of left- and right-handed leptons and right-handed neutrinos. } \label{tab:charge}
\end{table}
	
\subsection{Anomalies}
	
	It is important to study ACCs \cite{Allanach:2018vjg} even for low-energy theories which are  regarded as “only” Effective Field Theories (EFTs). In addition to cancellation of gauge and gravity anomalies,  the U(1)$^{\prime}$ charges of the fields must respect gauge invariance. By setting $Q_{H_u}+Q_{H_d}\neq 0$  the bare $\mu$ term is forbidden and gauge invariance of the superpotential implies
\begin{eqnarray}
	\label{eq:gauge_cond}
	0&=&Q_{S}+Q_{H_u}+Q_{H_d},
	 \\
	0&=&Q_{S}+Q_{D_x}+Q_{\overline{D}_x},
	 \\
	0&=&Q_{Q}+Q_{H_u}+Q_{U},
	 \\
	0&=&Q_{Q}+Q_{H_d}+Q_{D},
	 \\
	0&=&Q_{L}+Q_{H_d}+Q_{E},
	 \\	
	0&=&Q_{L}+Q_{H_u}+Q_{N}.
\end{eqnarray}
	In this work, we assume the charge of the exotic quark $Q_{D_x}$ to be the same as $Q_{H_u}$ for simplicity.  When non-universal charges are considered the above equations are subject to additional ramifications in accord with the Yukawa textures present in the theory.  For the model to be anomaly-free, the vanishing of
	U(1)$^{\prime}$-SU(3)$_C$-SU(3)$_C$,
	U(1)$^{\prime}$-SU(2)$_L$-SU(2)$_L$,
	U(1)$^{\prime}$-U(1)$_Y$-U(1)$_Y$,
	U(1)$^{\prime}$-graviton-graviton,
	U(1)$^{\prime}$-U(1)$^{\prime}$-U(1)$_Y$ and
	U(1)$^{\prime}$-U(1)$^{\prime}$-U(1)$^{\prime}$ anomalies should be satisfied, that is, the U(1)$^{\prime}$ charges of fields must
	obey the following ACCs:
\begin{eqnarray}
	0&=&3(2Q_{Q}+Q_{U}+Q_{D})+
	n_{D_x}(Q_{D_x}+Q_{{\overline{D}_x}}),
	\\
	0&=&3(3Q_{Q}+Q_{L})+Q_{H_d}+Q_{H_u},
	\\
	0&=&3(\frac{1}{6}Q_{Q}+\frac{1}{3}Q_{D}+
	\frac{4}{3}Q_{U}+
	\frac{1}{2}Q_{L}+Q_{E})
	+\frac{1}{2}(Q_{H_d}+Q_{H_u})\nonumber \\
	&+&3n_{D_x} Y^2_{D_x} (Q_{D_x}+
	Q_{{\overline{D}_x}}),
	\\
	0&=&3(6Q_{Q}+3Q_{U}+3Q_{D}+2Q_{L}+
	Q_{E}+Q_{N})
	+2Q_{H_d}+2Q_{H_u}\nonumber \\
	&+&Q_{S}+3 n_{D_x}
	(Q_{D_x}+Q_{{\overline{D}_x}}),
	\\
	0&=&3(Q^{2}_{Q}+Q^2_{D}-2Q^{
		2}_{U}-Q^{2}_{L}+Q^{ 2}_{E})-Q^{
		2}_{H_d}+Q^{2}_{H_u}+ 3n_{D_x} Y_{D_x}
	(Q^{2}_{D_x}- Q^{
		2}_{{\overline{D}_x}}),  \\
	0&=&3(6Q^{3}_{Q}+3Q^{ 3}_{D}+3Q^{
		3}_{U}+2Q^{ 3}_{L}+Q^{ 3}_{E}+Q^{
		3}_{N})+ 2Q^{ 3}_{H_d}+2Q^{
		3}_{H_u}+Q^{3}_{S}\nonumber \\
	&+& 3n_{D_x}(Q^{ 3}_{D_x}+Q^{
		3}_{{\overline{D}_x}}).
	\label{accs}
\end{eqnarray}
	
	From the above  equations one can easily extract the related conditions for non-universal charge selections affecting solely the leptonic sector.  As is well known,  quark charge deviations from universality may yield results with dangerously large Flavour Changing Flavour Current (FCNC) predictions. However, when non-universality is allowed only for leptons the problem is not that severe. Additionally, non-universal leptonic charges may bring desired predictions for recently discussed $B$-anomalies \cite{Frank:2019nwk,Frank:2020byg}.  
	
	By using the presented equations, it is easy to get family dependent conditions simply by replacing 
	$3 Q_{L}$ with $\Sigma_i Q_{L_i}$ terms  where  $L_i$ stands for any family of the leptons ($L,E,N$). 
	It is timely to now state that, in our construction, we have taken $n_{D_x} = 3$ colour triplet pairs with hypercharge $Y_{D_x}
	= -1/3$.  Also, it is appropriate to explicitly show our selections for the Yukawa couplings in the soft breaking terms in Eq. (\ref{UMSSM_NH}).
	
	
	As can be inferred from the ongoing discussion, while the quark sector is untouched, we demanded the masses of the electron and muon to come from non-holomorphic terms. With this choice, gauge invariance conditions are fulfilled as follows: 
\begin{eqnarray}
	0&=&Q_{L_1}+Q_{H_u}+Q_{E_1}\\
	0&=&Q_{L_2}+Q_{H_u}+Q_{E_2}\\
	0&=&Q_{L_3}+Q_{H_d}+Q_{E_3}.
\end{eqnarray}
	Similarly, for the neutrinos we resort to the following conditions:
\begin{eqnarray}
	0&=&Q_{L_1}+Q_{H_d}+Q_{N_1}\\
	0&=&Q_{L_2}+Q_{H_d}+Q_{N_2}\\
	0&=&Q_{L_3}+Q_{H_u}+Q_{N_3}.
\end{eqnarray}
	
	This setup enabled us to obtain numerical solutions for which we used a simple computer script yielding around 80 different solutions (the aforementioned sub-models) when $Q_{\rm Max}=10$. Of course, larger solution sets can be obtained when  $Q_{\rm Max}$ is increased. It is easy to obtain numerical solutions for instance even when $Q_{\rm Max}\sim10^2$. When $Q_{\rm Max}$ is large, actual physics does not differ since the extra gauge coupling is also scaled. However, extremely large values of $Q_{\rm Max}$ force the gauge coupling to be extremely small when they should not be smaller than the gravitational coupling. Besides this, when $Q$ is allowed to be very large, since solution sets are also enlarged, covering the emerging solutions (i.e., sub-models) can be problematic computationally. Hence, during numerical investigation we allowed the absolute value of any of the charges to be as large as 10 and scanned integer values satisfying all the equations of this subsection. Following this, we normalised  all the solutions to 1. This approach enabled us to keep the extra gauge coupling $g^\prime$  $\sim$ 0.41 at the weak scale. Our solutions are to be presented in visual format in the following sections.  
	
\subsection{The Exotic Sector}
	
	As mentioned, the presence of the exotics helps to satisfy ACCs. Beyond this, however, many theories predict their existence and their detection may be possible in the following years if they couple to known particles. Neither their existence nor their couplings are known yet, but we should study their properties under certain assumptions.
	In Eqs. (\ref{eq:superpot}) and (\ref{eq:soft}), $D_x$ and $ {\overline{D}_x} $ are exotic fermions with same charges as the down quarks. They are colour triplet and vector-like with respect to the MSSM, but chiral under the $U(1)^\prime$ symmetry. These exotics are similar to the ones discussed in the context of $E_6$ gauge groups \cite{Rosner:1999ub}  but they interact only with $ S $ and cannot mix with SM fermions in this model. The presence of the exotic fermions is required by ACCs but it is not obligatory as was shown for the minimal U(1)$^\prime$ extension in \cite{Demir:2005ti}. Alternatively, one can also assume the presence of the exotics such that they are very heavy and can be integrated out from the particle spectrum while their imprints should be preserved for ACCs. However, in this work, we assumed that the  exotic quarks exist in the particle spectrum and can be light.  As an experimental bound, they are expected to have masses larger than 1 TeV \cite{Kazana:2016goy}. We selected $ \kappa $ to be responsible for the masses of $D_x$ and $ {\overline{D}_x} $. Thus, the mass of the exotic fermion can be written as follows:
\begin{equation} 
	m_{D_x} =  
	\begin{array}{c}
	\frac{1}{\sqrt{2}} v_S \kappa. 
           \end{array}  
\end{equation}
	Furthermore, the mass squared matrix of the supersymmetric partners of the exotics can  be written as follows:
\begin{equation} 
	m^2_{\tilde{D}_x} = \left( 
	\begin{array}{cc}
	m_{\tilde{D}_x^L}^2 &\frac{1}{2} \Big(\sqrt{2} v_S T_{\kappa}  - v_d v_u \lambda \kappa \Big)\\ 
	\frac{1}{2} \Big(\sqrt{2} v_S T_{\kappa}  - v_d v_u \lambda \kappa \Big) &m_{\tilde{D}_x^R}^2\end{array} 
	\right),
\end{equation} 
	where
\begin{align} 
	m_{\tilde{D}_x^L}^2 &= \frac{1}{12} {\bf 1} \Big(6 g'^{2} Q_{D_x} \Big(Q_{H_d} v_{d}^{2}  + Q_{H_u} v_{u}^{2}  + Q_S v_{S}^{2} \Big) + g_{1}^{2} \Big(- v_{u}^{2}  + v_{d}^{2}\Big)\Big)  + \frac{1}{2}  \Big(2 m_{X}^2  + v_{S}^{2} {\kappa^2} \Big),\\ 
	m_{\tilde{D}_x^R}^2 &= \frac{1}{12} {\bf 1} \Big(6 g'^{2} Q_{\bar D_x} \Big(Q_{H_d} v_{d}^{2}  + Q_{H_u} v_{u}^{2}  + Q_S v_{S}^{2} \Big) + g_{1}^{2} \Big(- v_{d}^{2}  + v_{u}^{2}\Big)\Big)  + \frac{1}{2}  \Big(2 m_{\bar{X}}^2  + v_{S}^{2} {\kappa^2} \Big).
\end{align} 
	
\subsection{The $ Z^\prime $ Boson}
	
	Besides the singlet $S$ and its superpartner in the UMSSM,  the additional U(1)$^\prime$ symmetry introduces a new vector boson, $Z'$, and its supersymmetric partner, ${\tilde B}'$. After the breaking of the SU(2) $\times$ U(1)$_{Y}$ $\times$ U(1)$^\prime$ symmetry spontaneously, this new boson can mix with the $Z$ boson. The $Z-Z'$ mass matrix, that gives the physical mass eigenstates, is as follows:
\begin{eqnarray}
	\mathbf{M_{Z}^2} &=&
	\left(
	\begin{array}{cc}
	M_{ZZ}^2	&	M_{ZZ'}^2	\\
	M_{ZZ'}^2	&	M_{Z'Z'}^2	
	\end{array}
	\right)	= 
	\left(
	\begin{array}{cc}
	2g_{1}^2\sum_{i} t_{3i}^2\left| \left\langle \phi_{i}\right\rangle \right|^{2}	&	2g_{1}g^{\prime}\sum_{i} t_{3i} Q_{i}\left| \left\langle \phi_{i}\right\rangle \right|^{2}	\\
	2g_{1}g^{\prime}\sum_{i} t_{3i} Q_{i}\left| \left\langle \phi_{i}\right\rangle \right|^{2}	&	2g'^{2}\sum_{i} Q_{i}^2\left| \left\langle \phi_{i}\right\rangle \right|^{2}	
	\end{array}
	\right),
	\label{ZZpmatrix}
\end{eqnarray}
	where $t_{3i}$ is the weak isospin of the Higgs doublets or singlet while the $\left| \left\langle \phi_{i}\right\rangle \right|$'s stand for their VEVs. The matrix in Eq. (\ref{ZZpmatrix}) can be diagonalised by an orthogonal rotation and the mixing angle $\alpha_{ZZ'}$ can be written as 
\begin{eqnarray}
	\tan 2\alpha_{ZZ'} &=& \frac{2M_{ZZ'}^2}{M_{Z'Z'}^2-M_{ZZ}^2}.
	\label{az}
\end{eqnarray}
	The physical mass states of the $Z$ and $Z'$ are given by
\begin{eqnarray}
	M^2_{Z,Z'} &=&
	\frac{1}{2} 
	\left[
	M^2_{ZZ} + M^2_{Z'Z'} \mp 
	\sqrt{ \left( M^2_{ZZ} - M^2_{Z'Z'}\right)^2 + 4 M^4_{ZZ'}}
	\right] \,.
\end{eqnarray}
	
	The EW Precision Tests (EWPTs) put a strong bound on the $|\alpha_{ZZ'}| $ value to be less than a few times $ 10^{-3}$ \cite{Erler:2009jh}.  The most stringent lower bound on the $Z'$  mass has been set  by ATLAS in the di-lepton channel \cite{Aad:2019fac} while there is no specific bound on the mass of its supersymmetric partner $ {\tilde B}' $.
	Furthermore, the couplings of the $Z'$ boson to fermions are related to their currents described by Lagrangian and written as
\begin{eqnarray}
	{J^\prime}^\mu &= \sum_i \bar f_i \gamma^\mu[\epsilon_L^{{i}}P_L+\epsilon_R^{{\i}}P_R] f_i, \nonumber \\
	&= \frac{1}{2} \sum_i \bar f_i \gamma^\mu[g_v^{{i}}-g_a^{{i}}\gamma^5] f_i.
	\label{currents}
\end{eqnarray}
	where  $f_i$ is the field of the $i^{th}$ fermion and $\epsilon_{L,R}^{{i}}$ are the chiral couplings, which are the $U(1)^\prime$  charges of the left and right
	handed components of  fermion $f_i$, respectively. In addition, $g_{v,a}^{{i}} =\epsilon_L^{{i}}\pm\epsilon_R^{{i}}$
	are the corresponding vector and axial couplings \cite{Langacker:2008yv}. In writing Eq. (\ref{currents}), we neglect the mixing between the $Z$ and $ Z^\prime $ bosons because of the stringent experimental bound from EWPTs.
	
\subsection{The Neutralino Sector}
	
	Given the absence of any new charged gauged bosons in the UMSSM, the chargino sector remains the same as that in the MSSM. However, due to the mixing of $ {\tilde B}' $ and the fermionic partner of $ S $ with the MSSM gauginos and higgsinos, the UMSSM has a rich neutralino sector with six such states. Their masses and mixing can be given in the $(\tilde{B}',\tilde{B},\tilde{W},\tilde{h}_{u},\tilde{h}_{d},\tilde{S})$
	basis as follows:
\begin{equation}
	\mathcal{M}_{\tilde{\chi}^{0}}=\left(\begin{array}{cccccc}
	M_{1}' & 0 & 0 & g'Q_{H_{d}}v_{d} & g'Q_{H_{u}}v_{u} & g'Q_{S}v_S \\ 
	0 &M_{1}& 0&-\dfrac{1}{\sqrt{2}}g_{1}v_{d} & \dfrac{1}{\sqrt{2}}g_{1}v_{u}& 0 \\ 
	0 & 0&M_{2}&\dfrac{1}{\sqrt{2}}g_{2}v_{d}& -\dfrac{1}{\sqrt{2}}g_{2}v_{u}& 0 \\
	g'Q_{H_{d}}v_{d} & -\dfrac{1}{\sqrt{2}}g_{1}v_{d}&\dfrac{1}{\sqrt{2}}g_{2}v_{d}&0&-\dfrac{1}{\sqrt{2}}h_{s}v_{u}& -\dfrac{1}{\sqrt{2}}h_{s}v_{u} \\ 
	g'Q_{H_{u}}v_{u} &\dfrac{1}{\sqrt{2}}g_{1}v_{u}&-\dfrac{1}{\sqrt{2}}g_{2}v_{u}&-\dfrac{1}{\sqrt{2}}h_{s}v_S&0& -\dfrac{1}{\sqrt{2}}h_{s}v_{d} \\ 
	g'Q_{S}v_S &  0 & 0 & -\dfrac{1}{\sqrt{2}}h_{s}v_{u} & -\dfrac{1}{\sqrt{2}}h_{s}v_{d} & 0
    \end{array}
	\right),
\end{equation}
	where $M'_{1}$ is the mass of $\tilde{B}'$ after REWSB and the first row
	and column encode the mixing of $\tilde{B}'$ with the other
	neutralinos. Additionally, the sfermion mass sector also has extra contributions from the $D$-terms specific to the UMSSM. The diagonal terms of the sfermion mass matrix are modified by
\begin{eqnarray}
	\Delta_{\tilde{f}} 
	&=&	\dfrac{1}{2} g'Q_{\tilde{f}}(Q_{H_u}v_{u}^2+Q_{H_d}v_{d}^2+Q_{S}v_S^2),
	\label{deltaf}
\end{eqnarray}
	where $ \tilde{f} $ refers to sfermion flavours \cite{Sert:2010ma}.
	
	As can be seen from the given equations, in U(1)$^\prime$ models extra charges play a crucial role in determining the properties (and, hence, the manifestations) of the additional particles. In a top-down approach, one can take these charges to be given ($E_6$ models can be mentioned as an example within this context) and study phenomenology. Alternatively one can also follow a bottom-up approach and try to find preferred values by the charges from existing experimental data.  In the following section, we will follow the second procedure, which is therefore also serving the purpose of informing model building of a more fundamental theory embedding our UMSSM constructions, wherein U(1)$^\prime$ charges are predicted.
	
\section{Scanning Procedure and Experimental Constraints}
	\label{sec:scan}
	
	In our work, we have employed the $\spheno$ (version 4.0.0) package \cite{Porod:2003um} obtained with $\sa$ (version 4.14.3) \cite{Staub:2008uz,Staub:2010jh,Staub:2015kfa}. Through this package, all gauge and Yukawa couplings at the weak scale are evolved to the GUT scale that is assigned by the condition of gauge coupling unification, described as $ g_{1}=g_{2}=g' \approx g_{3}$. Here, $ g_{3} $ is allowed to have a small deviation from the absolute unification condition as it has the largest threshold corrections around the GUT scale \cite{Hisano:1992jj}. It is important to note that $\spheno$ guarantees the gauge coupling unification with this small deviations of $ g_3 $. Upon evaluating all model parameters after Spontaneous Symmetry Breaking (SSB)  along with gauge and Yukawa couplings back to the EW scale, the program  calculates the spectrum at the low scale for the boundary conditions given at $ M_{\rm GUT} $. These bottom-up and top-down processes are realised by running the Renormalisation Group Equations (RGEs). In the numerical analysis, we have performed random scans over the following parameter space of the UMSSM given in Tab.~\ref{paramSP}, 	where $ m_{0} $ is the universal  mass of the scalars while  $ M_{1/2} $ is the universal  mass of the gauginos at the GUT scale. Besides, $ T_0 $ is the trilinear coupling  and $ \tan\beta $ is the ratio of the VEVs of the MSSM Higgs doublets. Furthermore, $ T_{\lambda} $ and $T_{\kappa}$ are the SSB  strengths of the $ S H_u H_d $ and $ SD_x\bar{D_x} $ interactions. Finally, also  the 
  $\lambda$ and $\kappa$ couplings  as well as $v_S$, that is the VEV of $S$,  are varied.
\begin{table}[!t]
		\centering
		\setlength\tabcolsep{8pt}
		\renewcommand{\arraystretch}{1.4}
		\begin{tabular}{c|c||c|c}
			Parameter  & Scanned range & Parameter      & Scanned range \\
			\hline
			$m_0$ & $[0, 9]$ TeV     & $T_{\lambda}$ & $[-2.5, 2.5]$ TeV\\
			$M_{1/2}$        & $[0, 9]$ TeV & $T_{\kappa}$ & $[-2.5, 2.5]$ TeV\\
			$\tan\beta$ & $[1, 50]$  & $v_S$  & $[5, 45]$ TeV\\
			$\lambda$    & $[0.01, 0.4]$  & $T_{e}^\prime$ & $[-2, 2]$ TeV\\
			$\kappa$ & $[0.01, 0.8]$   & $T_{\mu}^\prime$ & $[-2, 2]$ TeV\\
			$T_0$    & $[-3 m_0, 3 m_0]$  &  \\
        \end{tabular}
		\caption{\sl\small Scanned parameter space of the UMSSM.}
		\label{paramSP}
\end{table}

	In the scanning of the UMSSM parameter space, the Metropolis-Hasting algorithm \cite{Belanger:2009ti} is used. After data collection, we implement Higgs boson and sparticle mass bounds \cite{Chatrchyan:2012ufa,Tanabashi:2018oca} as well as constraints from Branching Ratios (BRs) of $B$-decays such as $ {\rm BR}(B \rightarrow X_{s} \gamma) $ \cite{Amhis:2012bh}, $ {\rm BR}(B_s \rightarrow \mu^+ \mu^-) $ \cite{Aaij:2012nna} and $ {\rm BR}(B_u\rightarrow\tau \nu_{\tau}) $ \cite{Asner:2010qj}. We also require that the predicted relic density of the neutralino LSP agrees within 20\% (to conservatively allow for uncertainties on the  predictions)  with  the  recent Wilkinson Microwave Anisotropy Probe (WMAP) \cite{Hinshaw:2012aka} and Planck  results \cite{Ade:2013zuv,Aghanim:2018eyx}, $\Omega_{\rm CDM} h^2 =  0.12$.  The relic density of the LSP is calculated with $\mo$ (version 5.0.9) \cite{Belanger:2018ccd}. The experimental constraints can be summarised as follows\footnote{Here, $h$ is the  lightest CP-even MSSM Higgs state.}:
\begin{equation}
	\setstretch{1.8}
	\begin{array}{l}
	m_h  = 122-128~{\rm GeV},
	\\
	m_{\tilde{g}} \geq 2~{\rm TeV},
	\\
	0.8\times 10^{-9} \leq{\rm BR}(B_s \rightarrow \mu^+ \mu^-)
	\leq 6.2 \times10^{-9} \;(2\sigma~{\rm tolerance}),
	\\
	m_{\tilde{\chi}_{1}^{\pm}} \geq 103.5~{\rm GeV}, \\
	m_{\tilde{\tau}} \geq 105~{\rm GeV}, \\
	2.99 \times 10^{-4} \leq
	{\rm BR}(B \rightarrow X_{s} \gamma)
	\leq 3.87 \times 10^{-4} \; (2\sigma~{\rm tolerance}),
	\\
	0.15 \leq \dfrac{
		{\rm BR}(B_u\rightarrow\tau \nu_{\tau})_{\rm UMSSM}}
	{{\rm BR}(B_u\rightarrow \tau \nu_{\tau})_{\rm SM}}
	\leq 2.41 \; (3\sigma~{\rm tolerance}), \\
	0.0913 \leq \Omega_{{\rm CDM}}h^{2} \leq 0.1363~(5\sigma~{\rm tolerance}).
	\label{constraints}
	\end{array}
\end{equation}
	In addition to all constraints given in Eq. (\ref{constraints}), we have also applied the current  $Z'$  mass bounds from the $\sigma(pp \to Z'  \to ll)$ and $\sigma(pp \to Z' \to {\rm jets})$ channels, where $l=e,\mu$ and the hadronic final states do not include top (anti)quarks. The cross section values for the given processes at the LHC and a future $e^+e^-$ collider ave been calculated by using $\mg$ (version 2.6.6) \cite{Alwall:2014hca} along with the Leading Order (LO) set of NNPDF Parton Distribution Functions (PDFs) (version  2.3) \cite{Ball:2012cx}.

\section{Results}
\label{sec:results}
\begin{figure}[!t]
	\centering
	\includegraphics[scale=0.45]{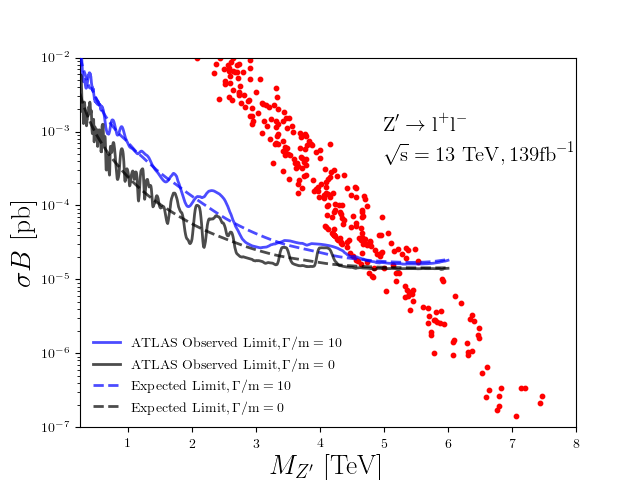}
	\includegraphics[scale=0.45]{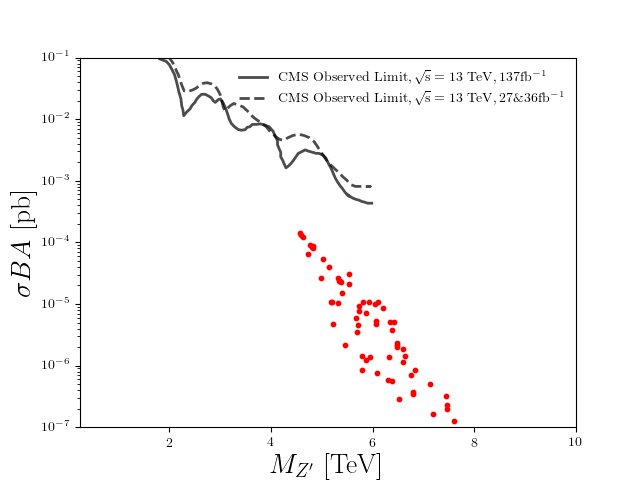}
	\caption{\sl\small The $Z'$ boson mass limits on $\sigma(pp \to $Z'$ \to ll)$ vs $M_{Z'}$ (left panel) and  $\sigma(pp \to $Z'$ \to {\rm jets})$ vs $M_{Z'}$ (right panel), where $ l $ describes electron and muon while jets refer to contributions from both the SM and  exotic quarks. All points plotted here satisfy all experimental constraints given in the previous section. In the right panel, we have additionally applied the $ Z^\prime $ mass constraint from the left panel. }
	\label{fig:z_prime_mass}
\end{figure}
In this section we present our results in the light of the experimental constraints from the previous section. First, let us focus on the mass limits of the $Z'$ boson from direct searches.
Fig. \ref{fig:z_prime_mass} shows the comparison of the experimental limits on the $Z'$ boson mass and its cross section as obtained from direct searches in the processes  $ pp \to ll $ at $ \mathcal{L}=137~{\rm{fb}}^{-1}$ \cite{Aad:2019fac} with full acceptance  and $ pp \to {\rm jets}$  with  experimental acceptance $A=0.5$ at ${\cal L}=137$ fb$^{-1}$ \cite{Sirunyan:2019vgj} as well as ${\cal L}= 27$ fb$^{-1}$ and $36$ fb$^{-1}$ \cite{Sirunyan:2018xlo}.  All points plotted here satisfy  all experimental constraints given in the previous section. According to our results, in the left panel, we find that the $Z'$ boson mass cannot be smaller than $ 4.5 $ TeV in the light of the ATLAS di-lepton results \cite{Aad:2019fac}. Furthermore, as can be seen from the right panel, the ATLAS results on the $Z'\to {\rm jets}$ channel   does not put any further limit on the $Z'$ mass, as in the right panel we have additionally applied the $ Z^\prime $ mass constraint from the left panel. It is important to note that our solutions also include contributions to the jet cross sections of the exotic quark $ D_x $  as well as SM quarks. In the reminder of this work, therefore, we use the $Z'$ boson mass allowed by all $Z'$ direct searches in the di-lepton channel as being $ M_{Z'} > 4.5  $ TeV\footnote{Note that the $\Gamma_{Z'}/M_{Z'}$ ratio for our  population of points is never larger than a few percent, so the factorisation procedure adopted by ATLAS and CMS in terms of cross section times BR ($\sigma B$) is applicable to our model.}.

\begin{figure}[!t]
	\centering
	\includegraphics[scale=0.45]{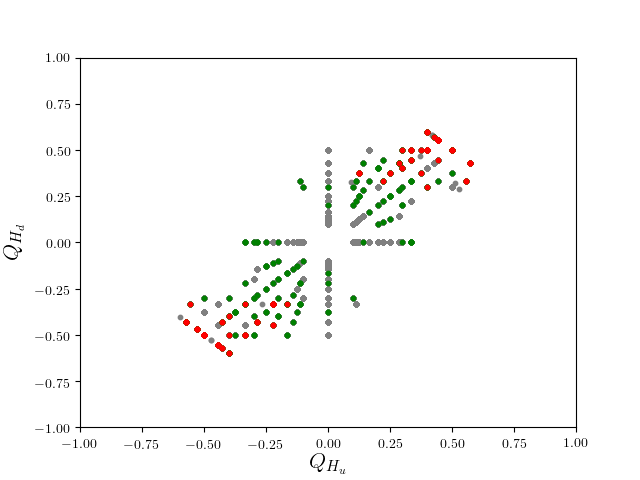}
	\includegraphics[scale=0.45]{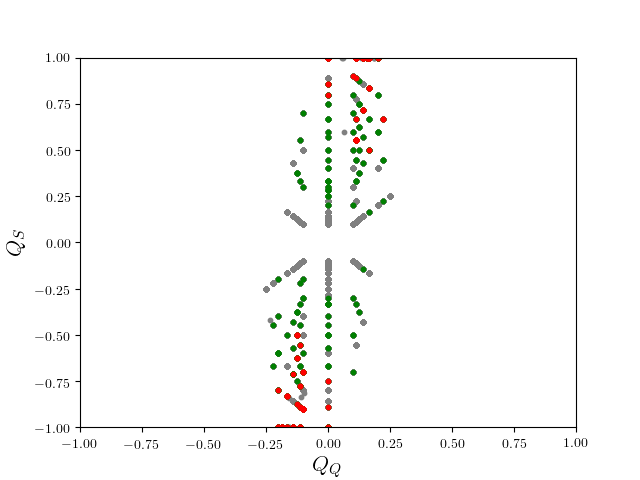}
	\includegraphics[scale=0.45]{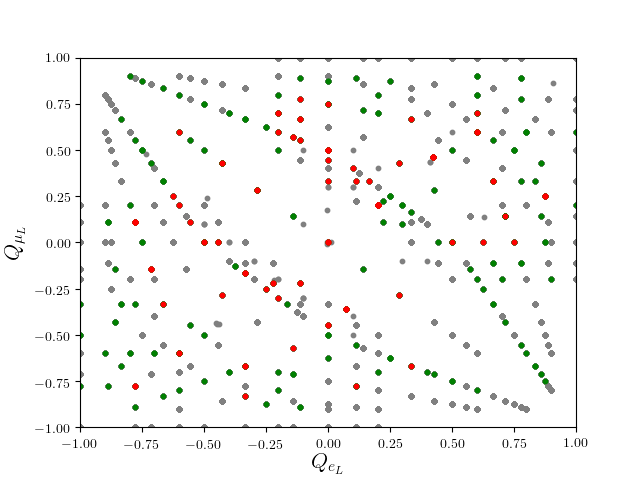}
	\includegraphics[scale=0.45]{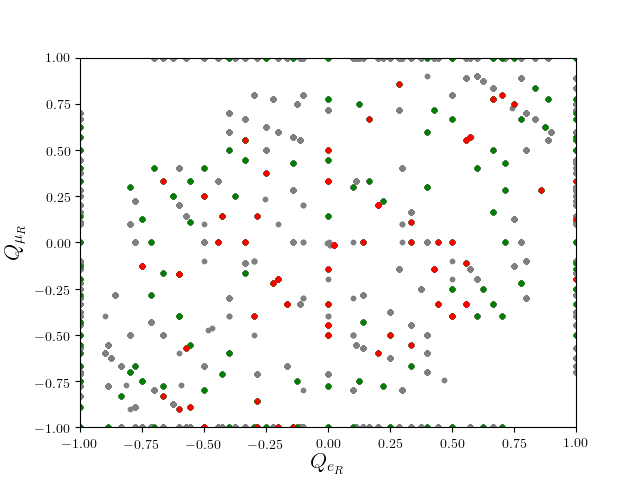}
	\caption{\sl\small The distributions of the U(1)$^\prime$ charges allowed by various theoretical and experimental conditions over the following planes: $ Q_{H_u} -Q_{H_d}$, $ Q_{Q} -Q_{S}$, $ Q_{e_L} -Q_{\mu_L}$ and $ Q_{e_R} -Q_{\mu_R}$. All points are consistent with REWSB, ACCs and neutralino LSP. Green points are a subset of the gray ones as  they also satisfy  all experimental constraints from Section \ref{sec:scan}. Red points are a subset of the green ones as  they are also compatible with the $ Z^\prime $ boson mass bounds in Fig. \ref{fig:z_prime_mass}.}
	\label{fig:charges}
\end{figure}
Fig. \ref{fig:charges} depicts the  U(1)$^\prime$ charge sets satisfying various theoretical and experimental bounds. The gray points correspond to configurations compliant with ACCs, REWSB and a neutralino as the Lightest Supersymmetric Particle (LSP). The green points form a subset of the gray ones as they also satisfy all the experimental constraints from Section \ref{sec:scan}, i.e., the exotic mass limits, the constraints from the rare $B$-decays and the bound on the relic abundance of the neutralino LSP. Red points instead form a subset of the green ones as they are also compatible with the $ Z^\prime $ boson mass bounds from direct searches shown in Fig. \ref{fig:z_prime_mass}. As can be seen from the top right panel, while ACCs allow a relatively large solution set, when all constraints are applied,  charges are restricted to certain regions, e.g.,  the $ -0.5 \lesssim Q_{H_u}, Q_{H_d} \lesssim 0.5$ domain is favoured but there is no solution in which these two charges may have opposite signs. Also notice that these charges are always far away from zero (for both of the symmetric solutions) since  $Q_S=-(Q_{H_u}+Q_{H_d})$. Furthermore, it can  easily be read, from the top right panel, that the lower limit on $ |Q_S| $ should be  $\sim 0.4$ after applying  all theoretical conditions and experimental constraints. It is also important to note that ACCs allow the $ Q_Q $ charge to be in a narrow interval, as $ -0.25 \lesssim Q_Q \lesssim 0.25 $. In contrast, according to the bottom panels, leptonic charges are allowed to take a wide span of  different values. As we expected, in  all panels of this figure, the charge patterns corresponding to the gray points  are symmetric, while those  also  corresponding to the green and red ones are not. It is then clear that this symmetry at the theoretical level is lost when applying  constraints from experimental observables, unsurprisingly, as interference effects are not negligible in many of these. This will have bearings on the construction of the fundamental theory behind the UMSSM wherein charges are predicted. 

As a next step, we studied  the $ Z^\prime $ decay modes in presence of exotic quarks lighter than $M_{Z'}/2$ and 
  non-universal leptonic charges. In Fig. \ref{fig:z_prime_br} we present BRs of the $Z'$ for different decay channels, BR$(Z' \to XX) $, where 
$XX=WW, ll$ (now including tauons), $qq$ (now including top (anti)quarks), $\nu\nu$ and $D_xD_x$  (left panel), and individual leptonic channels, BR$(Z' \to ll) $,  where $l=e,\mu,\tau$ (right panel), as a function of $M_{Z'}$. Our colour convention can be read from the figure. As can be seen from the left panel, BRs of the decays into exotic quarks can be as large as $\sim$ 30 $\%$ in our U(1)$^\prime$ model, a fact that should not be overlooked in future searches for $Z'$ states. Similarly, when non-universal charges are significantly different from each other, $Z'$ decay rates into electrons,  muons and  tauons  may in turn be different as can be seen from the right panel, which 
may also have important experimental implications in not only $Z'$ boson searches but also generic di-lepton measurements. We will dwell on these two aspects later on.

\begin{figure}[!t]
	\centering
	\includegraphics[scale=0.45]{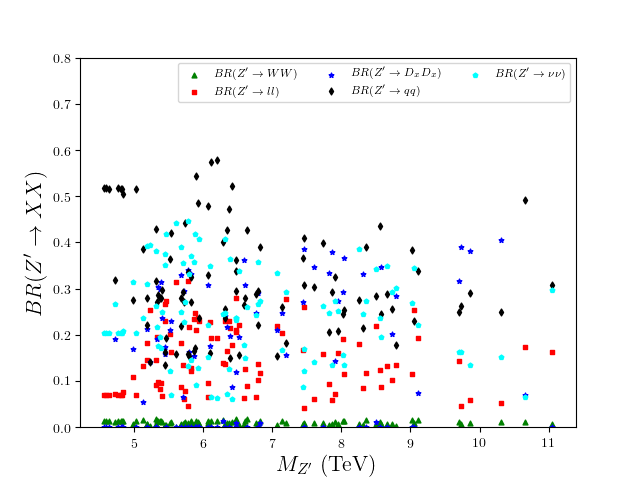}
	\includegraphics[scale=0.45]{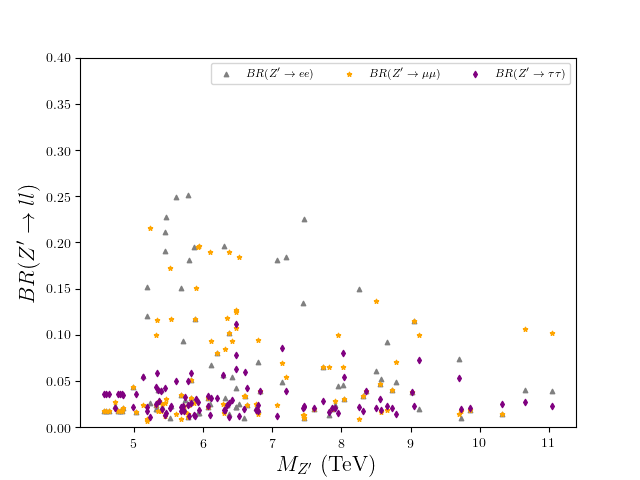}
	\caption{\sl\small The BRs of the $Z'$ for different decay channels, BR$(Z' \to XX) $ as a function on $M_{Z'}$ (left panel) and BR$(Z' \to ll) $ as a function on $M_{Z'}$ (right panel), where $XX$ represents all two-body final states while $ll$ describes individual leptonic final states.}
	\label{fig:z_prime_br}
\end{figure}

\begin{figure}[!t]
	\centering
	\includegraphics[scale=0.45]{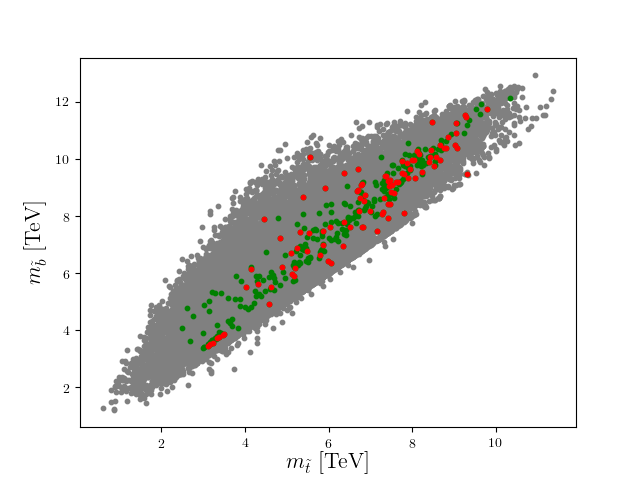}
	\includegraphics[scale=0.45]{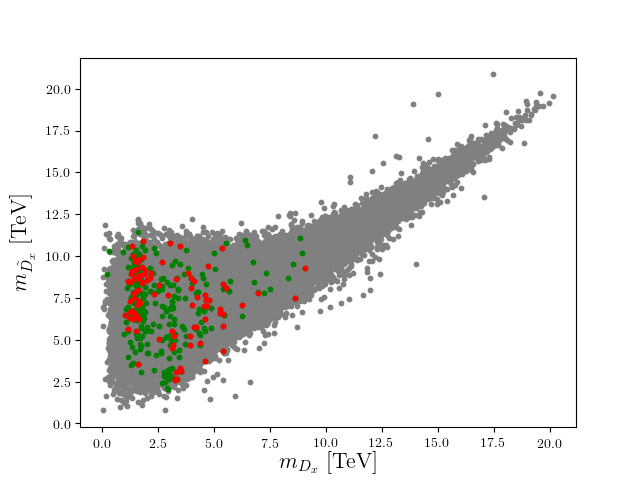}
	\includegraphics[scale=0.45]{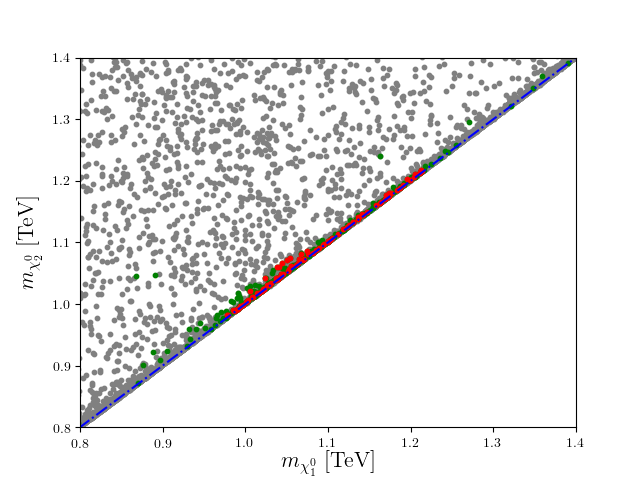}
	\includegraphics[scale=0.45]{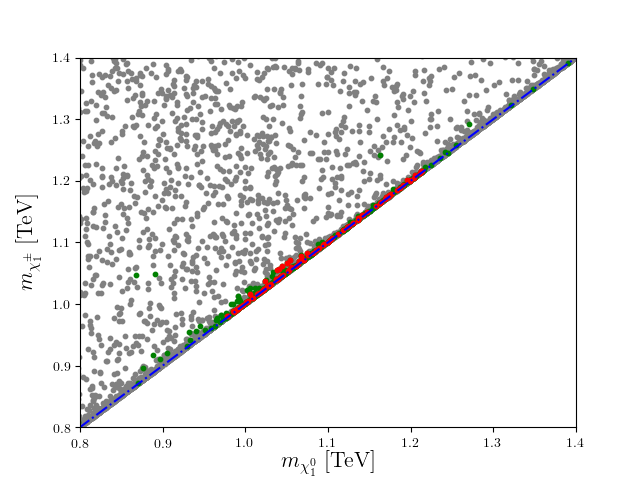}
	\caption{\sl\small The mass spectrum of the lightest chargino, two lightest neutralinos, stops, sbottoms and exotic (both fermionic and scalar) states over the following planes: $ m_{\tilde{t}}-m_{\tilde{b}} $, $ m_{D_x}-m_{\tilde{D_x}} $, $ m_{\tilde{\chi}_{1}^{0}}-m_{\tilde{\chi}_{2}^{0}} $ and $ m_{\tilde{\chi}_{1}^{0}}-m_{\tilde{\chi}_{1}^{\pm}} $. The colour convention  is the same as in Fig. \ref{fig:charges}. }
	\label{fig:spectrum}
\end{figure}

  A portion of the particle mass spectrum is presented in Fig \ref{fig:spectrum}. As can be seen from the top left panel, stop and sbottom masses are heavy in general and one can extract a $\sim$ 3 TeV lower mass limit. In contrast, the mass of the exotic quarks can be as light as 1.1 TeV, while the exotic scalar masses should be larger than 2.5 TeV. According to the bottom panels, the LSP, i.e., the neutralino DM, is relatively heavy as $ 1$ TeV $\lesssim m_{\tilde{\chi}_{1}^{0}} \lesssim 1.3 $ TeV. Furthermore, due to relic density constraints, the lightest neutralino and lightest chargino are very degenerate in mass. Such solutions can be seen for Higgsino-like LSP cases where not only the lightest neutralino but also the lightest chargino get their masses from Higgsinos. These solutions favour the chargino-neutralino coannihilation channels which reduce the relic abundance of the LSP,  such that the latter can be consistent with  experimental bounds. {In addition to these, since it has a potential to put constraints on the leptonic non-universal charges, we also looked for the $ R = BR(B \rightarrow s\ \mu^{+} \mu^{-} ) / BR(B \rightarrow s\ e^{+}  e^{-} )$ predictions of our model. For this aim, we used the filtered points passing all tests and we observed that our models R prediction does not deviate from the SM one sensibly, differing by $3-4\%$ at the most.}

\subsection{Collider Signatures}

We now investigate the observability of the exotic quarks and non-universality of the $ U(1)^\prime $ charges at  present and future colliders. As we will see,  the $ Z^\prime $ decays are important to facilitate observation of both sectors. Exotic quarks mainly interact with gluons and $ Z^\prime $ at the tree level while their couplings to the $ Z $ boson is extremely suppressed by the $ Z-Z^\prime $ mixing angle. Furthermore, due to the non-universal leptonic charges, the cross sections of  lepton pair production  for each flavour may differ, owing to the different $ Z^\prime $ mediation.
First, we start with hadron colliders. Then, we move on to lepton machines.
\begin{figure}[!t]
	\centering
	\includegraphics[scale=0.4]{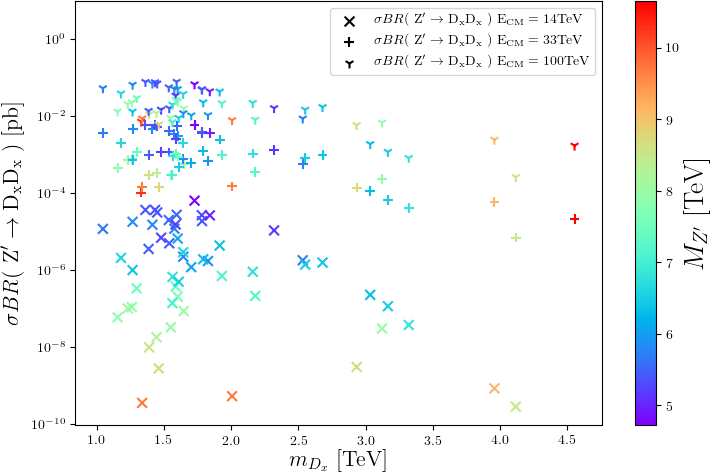}
	\includegraphics[scale=0.45]{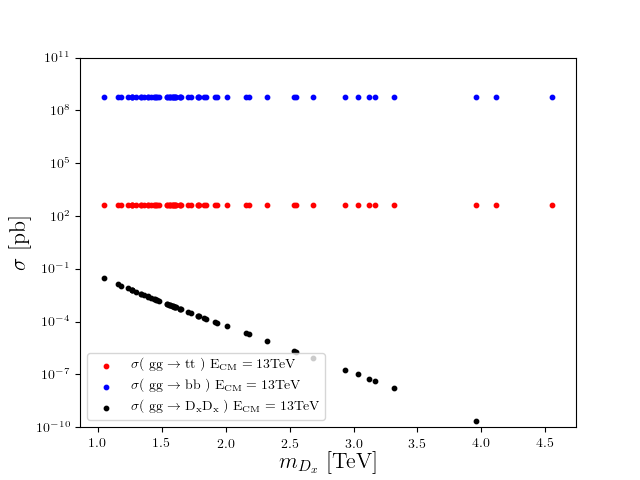}
	\caption{\sl\small The pair production cross sections of all exotic quarks through $Z'$ mediation {in terms of relevant $ Z^\prime $ mass} (left panel) at the HL-LHC (14 TeV), HE-LHC (33 TeV) and VLHC/FCC-hh (100 TeV) and those mediated by QCD (right panel), the latter alongside those for top and bottom quarks at 13 TeV. }
	\label{fig:exotic_LHC}
\end{figure}

{Fig. \ref{fig:exotic_LHC} shows the pair production cross sections of exotic quarks $ D_x $ through the $Z'$ (left panel)  at the High-Luminosity LHC (HL-LHC) \cite{Gianotti:2002xx,Apollinari:2015bam}, with $\sqrt s=$ 14 TeV, High-Energy LHC (HE-LHC) \cite{Abada:2019ono}, with $\sqrt s=$  33 TeV,  and Very LHC (VLHC)/Future Circular Collider in $pp$ mode (FCC-hh) \cite{Mangano:2017tke, Abada:2019lih}, with $\sqrt s=$ 100 TeV. Among the red points presented in Fig. \ref{fig:charges} we filtered the relevant points allowing $Z'$ decays to exotic quark pairs only {while the colour bar displays the mass of the relevant $ Z^\prime $ boson.} (Notice that, the same points are also used in Fig. \ref{fig:non_uni_LHC} below.)  Clearly, given the luminosity foreseen at the three machines, of 3, 15 and 30 ab${^{-1}}$, respectively, the (resonant) process $pp\to Z'\to {D_x}{D_x}$, carrying information about the underlying structure of the UMSSM, 
 can produce a sizable number of events. However, these will be overwhelmed by not only   ${D_x}{D_x}$ production via QCD but also $bb$ and $tt$ events from the SM, as all such processes would produce similar hadronic final states. This is made clear by the right panel of the figure, e.g., at the current LHC energy of 13 TeV.  The colour coding identifying the various processes can be read from the legend of the panels.}

\begin{figure}[!t]
	\centering
	\includegraphics[scale=0.45]{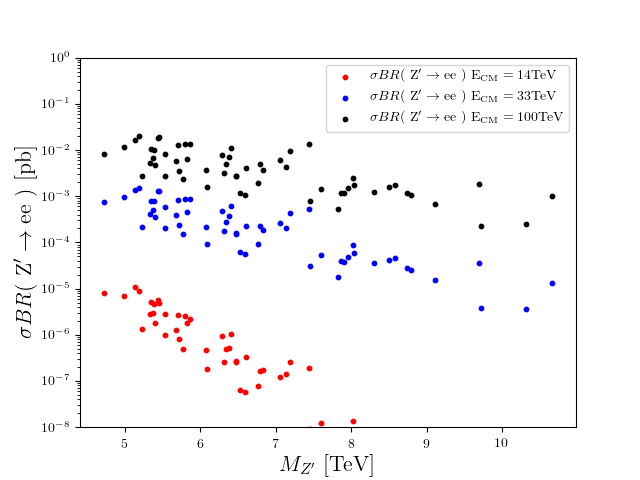}
	\includegraphics[scale=0.45]{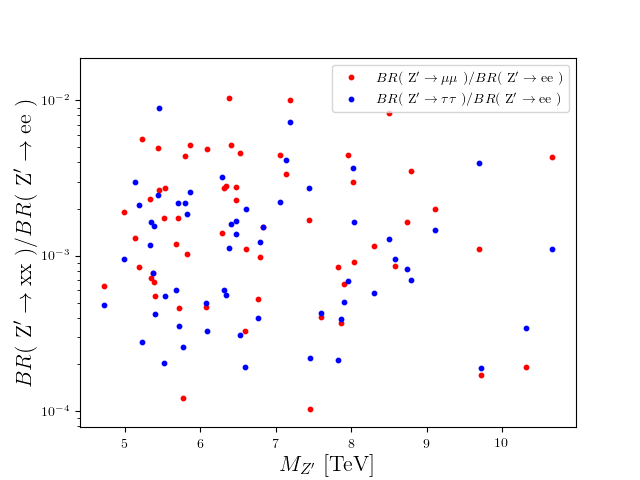}
	\caption{\sl\small The production cross sections of  electron pairs (left panel) through the $Z'$ at the HL-LHC, HE-LHC  and VLHC/FCC-hh and the ratio  of the BRs of $Z' \to \mu \mu $ and $Z'\to \tau \tau $ relative to that of  $Z' \to e e $ (right panel).}
	\label{fig:non_uni_LHC}
\end{figure}

In Fig. \ref{fig:non_uni_LHC}, we probe the observability of flavour non-universality at, again, the HL-LHC, HE-LHC  and VLHC/FCC-hh. The panel on the left depicts signal cross sections of electron positron pair creation via a  $Z'$. The panel on the right show muon ($x=\mu$) and tauon ($x=\tau$) BRs divided by the electron ($e$) one, from which one can infer that the BRs into muons and tauons can be 10 times larger or smaller than that into electrons. While this prediction is very significant, due to overall value of the signal cross sections, it is not very probable to see such a difference at 14  or 33 TeV, while there could be some chances of probing it at 100 TeV, given the aforementioned values of luminosity at these machines. Again, the colour coding identifying the various processes 
can again be read from the legend of the panels. Also, the population of points used in this figure is the same as in the previous one.

\begin{table}[!t]
	\renewcommand{\arraystretch}{1.3}\setlength\tabcolsep{6pt}
	\begin{center}
		\small
		\begin{tabular}{c||c|c}
		  Parameters &  {BP-I} & {BP-II}  \\ \hline
		  $ Q_{Q}  $  & 0.156                 &  $-0.125$                   \\ \hline
		  $ Q_{U}  $  & 0.373               & 0                    \\ \hline
		  $ Q_{D}  $  & 0.313                 & $-0.25$                        \\ \hline
		  $ Q_{L_1}  $  & 0.071                  & 0.875                         \\ \hline
		  $ Q_{L_2}  $  &  $-0.36$                 & 0.25                   \\ \hline
		  $ Q_{L_3}  $  & $-0.12$	                 & $-0.5$            \\ \hline	
		  $ Q_{N_1}  $  & 0.45                  & $-1$                           \\ \hline		
		  $ Q_{N_2}  $   & 0.89                 & $-0.375$            \\ \hline	
		  $ Q_{N_3}  $   & 0.65                 & 0.375                  \\ \hline
		  $ Q_{E_1}  $  & $-0.6$                  & $-0.75$                            \\ \hline		
		  $ Q_{E_2}  $   & $-0.17$                 & $-0.125$            \\ \hline	
		  $ Q_{E_3}  $   & 0.59                 & 0.125                  \\ \hline	
		  $ Q_{H_u}  $   &   $-0.53$          	 & 0.125                 \\ \hline 	
		  $ Q_{H_d}  $   &   $-0.47$           	 & 0.375                 \\ \hline 
		  $ Q_{S}  $   & 1                 & $-0.5$                 \\ \hline	
		  $ Q_{{D_x}}  $   &   $-0.53$          	 & 0.125                 \\ \hline 	
		  $ Q_{{{\overline{D}}_x}}  $   &   $-0.47$           	 & 0.375                 \\ \hline 
		   \hline												
			$M_{Z'}$                 &  5388 GeV  & 5452 GeV  \\ \hline 
			$m_{D_x}$	       &  1362 GeV & 1386 GeV  \\
		\end{tabular}
		\caption{\sl\small U(1)$^\prime$ charges, $Z'$ and exotic quark masses for  two Benchmark Points (BPs) of the UMSSM: {BP-I} and {BP-II}.}
		\label{tab:charge_BM}
	\end{center}
\end{table}

Next, we investigate both sectors, the exotic and leptonic one, of the UMSSM at the Compact Linear Collider (CLiC), which is a concept linear collider for electron-positron collisions \cite{Aicheler:2012bya, Aicheler:2019dhf}. This is proposed to be built at CERN and foreseen to reach up to 3 TeV in  Centre-of-Mass  energy ($E_{CM}$)  in the final stage of operations \cite{Charles:2018vfv}. In this analysis, to illustrate the scope of this $e^+e^-$ machine, we use two BPs having similar $ Z^\prime $ and exotic masses but different  U(1)$^\prime $ charge configurations, as shown in Tab. \ref{tab:charge_BM}. Even if the direct observation of any potential $Z'$ boson is not possible at CLiC, due to its lower mass bound from the LHC, in the suitable kinematical conditions, its impact on both quark and lepton pair production through electron-positron collisions can be detected indirectly.

 In Fig. \ref{fig:CLiC1}, we show the cross section for quark pair production (the leading contribution to $ e^+e^- \rightarrow$ jets) versus the CLiC collision energy with three different model configurations,
 SM, SM + $ Z^\prime $ and SM + $ Z^\prime $ + exotic quarks, for BP-I (left panel) and BP-II (right panel). For both BPs, it is clear that, as soon as $ E_{CM}\approx 2 m_{D_x} $, the hadronic final state always sees a significant increase of the cross section, well beyond the expected theoretical and experimental uncertainties. Furthermore, for {BP-I}, this happens without any significant contributions of the $Z'$ (as the SM and SM + $Z'$ cross sections coincide throughout), while for {BP-II} the situation is opposite (since the SM and SM + $Z'$ cross sections differ throughout). Quite
remarkably, in the latter case, the $Z'$ presence manifests itself via a large negative interference between the  $ \gamma, Z $ and $ Z^\prime $ channels, as can be expected by inspecting the  charge configurations given  in Tab. \ref{tab:charge_BM}. 

\begin{figure}[!t]
	\centering
	\includegraphics[scale=0.45]{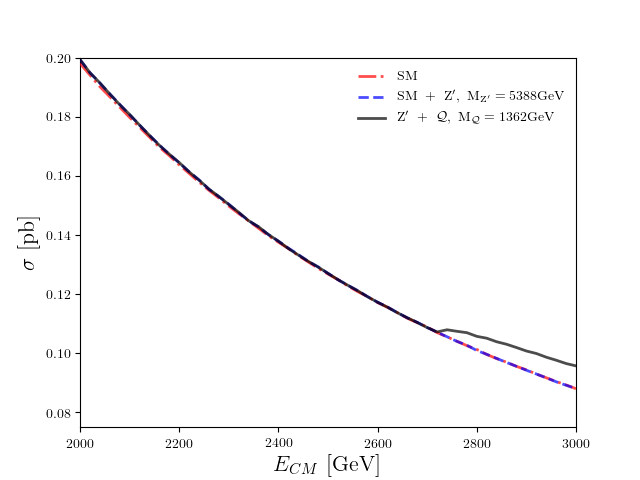}
	\includegraphics[scale=0.45]{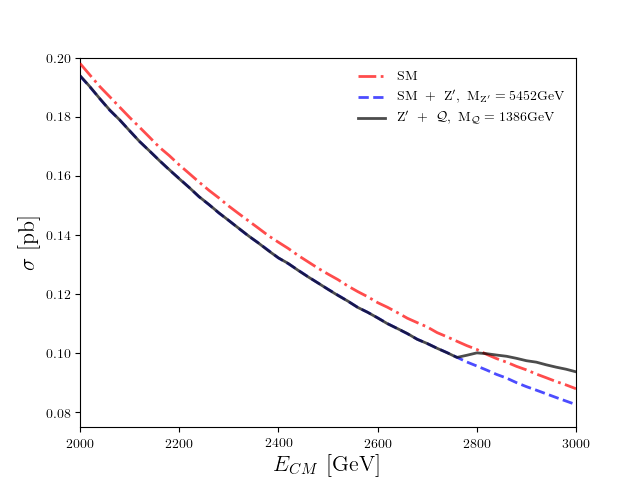}
	\caption{\sl\small The predicted cross section for quark pair production versus collision energy at CLiC in the case  of  the SM, SM + $Z'$  and  SM + $Z'$ + exotic quarks for BP-I (left) and BP-II (right).}
	\label{fig:CLiC1}
\end{figure}

\begin{table}[!h]
	\renewcommand{\arraystretch}{1.3}\setlength\tabcolsep{6pt}
	\begin{center}
\hspace*{-0.5cm}
		\begin{tabular}{c|c c c c c c c c c c c c}
			& $g^{e}_v$ & $g^{e}_a$ & $g^{\mu}_v$ & $g^{\mu}_a$
			& $g^{\tau}_v$  & $g^{\tau}_a$ & ${\cal{F}}_{ee}$ & ${\cal{G}}_{ee}$& ${\cal{F}}_{\mu\mu}$ & ${\cal{G}}_{\mu\mu}$& ${\cal{F}}_{\tau\tau}$ & ${\cal{G}}_{\tau\tau}$\\
			\hline\hline
			{BP-I}  & $-0.53$ & 0.67 & $-0.53$ &  $-0.19$ & 0.46 & $-0.71$ & 0.53 & 0.50 & 0.23 & $-0.14$  & 0.53 & 0.47 \\
			{BP-II} & 0.125 & 1.625 & 0.125 & 0.375 & $-0.375$ & $-0.625$ & 7.05  & 0.16 & 0.41 & 0.038  & 1.41 & 0.19 \\									
		\end{tabular}		
		\caption{\sl\small The vector and axial couplings of the $ Z^\prime $ boson to leptons with family non-universal $  U(1)^\prime $ charges for {BP-I} and {BP-II} together with the values of the ${{\cal{F}}_{ll}}$ and ${{\cal{G}}_{ll}}$ functions introduced in Eq. (\ref{eq:F_G}). }
		\label{tab:benchmarks_couplings}
	\end{center}
\end{table}

Additionally, the non-universality of the leptonic charges mainly affects the $ Z^\prime $ decay modes to each lepton flavour as discussed in Section \ref{sec:results}. So, the $ Z^\prime $ contribution to lepton pair production, $ ff \rightarrow ll $, is expected to be different for each family. Let us analyse this contribution. The diagonal amplitude-squared term of the lepton pair production through the $ Z^\prime $ boson can be written as 
\begin{eqnarray}
\,|{\cal{M}}_i\left(f\, \overline{f} \rightarrow Z' \rightarrow \ell^+
\ell^-\right)|^2=\, F(s; v,a)\,,
+ \,G(s; v,a)
\end{eqnarray}
where $F(s; v,a)$ and $G(s; v,a)$ are given by \cite{delAguila:1986klm}
\begin{eqnarray}
F(s;
v,a)\approx\frac{(g_{v}^{f}\,g_{v}^{f}+g_{a}^{f}\,g_{a}^{f})\,
	(g_{v}^{l}\,g_{v}^{l}+g_{a}^{l}\,g_{a}^{l})}{(s-M_{Z'}^2+iM_{Z'}\Gamma_{Z'})^2},
\end{eqnarray}
\begin{eqnarray}
G(s;
v,a)\approx\frac{(g_{v}^{f}\,g_{a}^{f}+
	g_{v}^{f}\,g_{a}^{f})\,(g_{v}^{l}\,g_{a}^{l}+g_{v}^{l}\,g_{a}^{l})}
{(s-M_{Z'}^2+iM_{Z'}\Gamma_{Z'})^2}\,,
\end{eqnarray}
where $ g_v^{f,l} $ and $ g_a^{f,l} $ are the vector and axial couplings of the $ Z^\prime $ to the initial and final state fermions, which are dependent upon the U(1)$^\prime $ charges, as shown in Eq. (\ref{currents}). Here, $\sqrt s=E_{CM}$ and $ \Gamma_{Z'} $ is the total decay width of the  $ Z^\prime $ boson. So, we can introduce two parameters to which the diagonal term of the amplitude-squared for $ ee \rightarrow ll $ is proportional, as follows:
\begin{eqnarray}
\label{eq:F_G}
{\cal{F}}_{ll}&=&({g^{e}_v}{g^{e}_v}+{g^{e}_a}{g^{e}_a})({g^{l}_v}{g^{l}_v}+{g^{l}_a}{g^{l}_a}),
\nonumber \\	
{\cal{G}}_{ll}&=&({g^{e}_v}{g^{e}_a}+{g^{e}_v}{g^{e}_a})(({g^{l}_v}{g^{l}_a}+{g^{l}_v}{g^{l}_a}).
\end{eqnarray}
The values of the $ g_v^l $, $ g_a^l $ couplings as well as the $ {\cal{F}}_{ll} $ and $  {\cal{G}}_{ll} $ functions  for BP-I and BP-II are given in 
Tab.~\ref{tab:benchmarks_couplings}. Fig.~\ref{fig:CLiC2} shows the ratio of the  lepton pair production cross section predicted in the UMSSM with respect to the  SM one for electron, muon and tauon pair production as a function of the collision energy at CLiC for BP-I (left panel) and BP-II (right panel). As can be seen from the figure, the differences amongst  the normalised cross sections for $ee$, $\mu\mu$ and  $\tau\tau$ production can be very significant, again, beyond theoretical and experimental uncertainties. 
Like in the case of the hadronic cross sections, also for the one of the leptonic ones, the different trends can be ascribed to the different leptonic charges entering the $Z'$ and (marginally, through the $Z-Z'$ mixing) $Z$ contributions.

\begin{figure}[!t]
	\centering
	\includegraphics[scale=0.45]{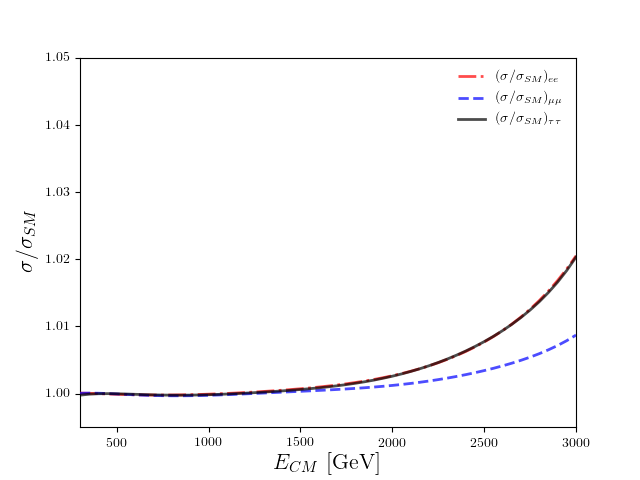}
	\includegraphics[scale=0.45]{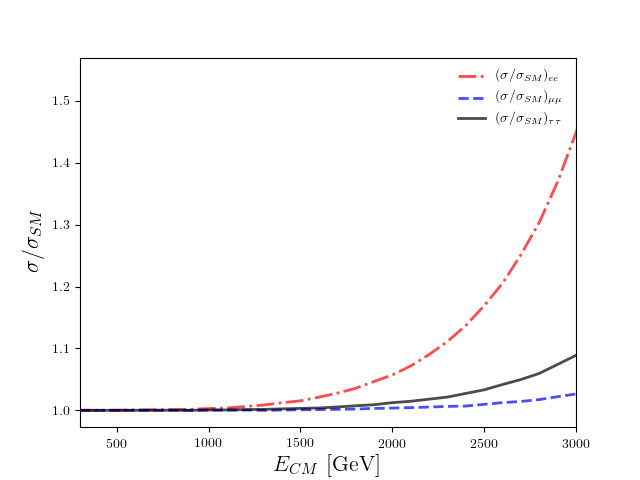}
	\caption{\sl\small The ratio of the predicted cross section for electron, muon and tauon pair production in the UMSSM relative to the SM values as a function of the collision energy at CLiC for BP-I (left) and BP-II (right).}
	\label{fig:CLiC2}
\end{figure}

\section{Conclusions}
\label{sec:cons}

In this paper, we have  studied potential effects of family non-universal U(1)$^\prime$ charges emerging in an UMSSM containing exotic coloured states, 
  fermionic and scalar, onto observables at colliders, both present and future ones. We have done so after implementing both theoretical and experimental constraints. The former include the request of satisfying ACCs, generating successful REWSB and yielding a neutralino as the LSP. The latter include
relic density, $B$-physics and collider bounds. While the theoretical conditions imposed allows for a symmetric distribution of such charges, wherein
$Q_{H_u}\sim Q_{H_d}$ is preferred  and $Q_S$ cannot be very small (these are the doublet and singlet charges),  the enforcement of the experimental 
ones renders the eventual solutions significantly different with a noticeable  loss of symmetry. Nonetheless, the surviving charges  produce UMSSM configurations inducing phenomenological manifestations that could be probed at high energy accelerators  presently discussed as successors to the LHC.

These include both hadronic (HL-LHC, HE-LHC and VLH/FCC-hh) and leptonic (CLiC) machines and the smoking-gun signatures to be pursued are final states capturing either the presence of the exotic states (i.e., hadronic ones) or that of electron, muons and tauons (i.e., leptonic ones) in proportions different from those predicted by the SM, possibly exploiting $Z'$ mediation, as interactions of the new U(1)$^\prime$ gauge bosons with such objects would carry key information about the underlying structure of the UMSSM. Following our numerical investigations, we have shown that a future $e^+e^-$ machine operating well beyond the TeV scale would be the ideal laboratory almost free from backgrounds to test this dynamics while future $pp$ ones would suffer from limitations connected to either small signal cross sections or overwhelming QCD noise. In practice, the study of the processes $e^+e^-\to $ hadrons and $e^+e^-\to f\bar f$, where $f=e,\mu,\tau$, could result in the possibility of extracting the parameters of both the exotic and leptonic sector so as to guide the formulation of the true theory yielding the UMSSM as its low scale manifestation.

\section{Acknowledgments}

SM is supported  in part through the NExT Institute and the STFC Consolidated
Grant No. ST/L000296/1.  The work of YH is supported by The Scientific and Technological Research Council of Turkey (TUBITAK) in the framework of  2219-International Postdoctoral Research Fellowship Program. We would like to thank Subhadeep Mondal for providing an example code to produce SARAH model files. The authors also acknowledge the use of the IRIDIS High Performance Computing Facility and associated support services at the University of Southampton in the completion of this work.

%






\bibliography{NUMSSM.bib}

\end{document}